\documentstyle[preprint,aps,epsfig]{revtex}

\begin{document}
\preprint{\font\fortssbx=cmssbx10 scaled \magstep2
\hfill$\vcenter{\hbox{\bf CERN-TH/95-262}
}$}
\title{Stimulated Neutrino Conversion in the  CERN Beam}
\author{M.\ C.\ Gonzalez-Garcia}
\address{Theory Division,   CERN,
CH-1211 Geneva 23, Switzerland.}
\author{F. Vannucci}
\address{LPNHE, Universit\'e Paris VII, F-75251 Paris CEDEX 05, France.}
\author{J.\ Castromonte}
\address{Departamento de F\'{\i}sica y Matem\'aticas,\\
Universidad Peruana Cayetano Heredia, Lima, Per\'u.}

\maketitle
\thispagestyle{empty}
\begin{abstract}
We discuss the possibility of searching for anomalous magnetic transitions of
neutrinos in the CERN beam induced by the absorption or emission of  low-energy
photons in a high-quality resonant cavity such as the LEP  radio-frequency
cavities. With the attainable sensitivities of the present CERN neutrino
detectors, this experiment would impose strong limits on this transition and on
the  radiative decay lifetime of neutrinos with masses in the range of interest
to the resolution of the dark matter  solar and atmospheric neutrino puzzles.
\\
\noindent
{\bf CERN-TH/95-262} \\
\noindent
{\bf October 1995}
\end{abstract}
\newpage
The existing indications for non-zero neutrino masses include the deficit of
solar electron neutrinos \cite {solar},  the deficit in the  atmospheric muon
neutrino flux \cite{atmos} and the indications favouring a hot component in the
dark
matter of the Universe \cite{dark1,dark2}. All solar neutrino experiments
\cite{solar} find fewer $\nu_e$'s than predicted theoretically. If the current
solar models are correct, the explanation
for this deficit relies on the oscillation of $\nu_e$ to another neutrino
species with  a mass difference $ \Delta m^2_{ei}\approx 10^{-5}$ eV$^2$.  Some
atmospheric neutrino experiments \cite{atmos} also observe a deficit in  the
ratio of experimental-to-expected ratio of muon-like to electron-like events.
An explanation for this anomaly relies on  $\nu_\mu$
oscillations into another flavour with a mass difference  $\Delta m^2_{\mu
i}\approx  10^{-2}$--$ 10^{-3}$ eV$^2$.

Massive neutrinos also play an important role in cosmology as the substance of
hot dark matter. Currently, the cosmological best-fit scenario includes a
mixture of cold plus hot dark matter \cite{dark1}. This
translates into an upper limit on neutrino masses \cite{dark2}: ${\displaystyle
\sum_i} m_{\nu_i}\lesssim 10 \; \mbox{eV}$.
Putting all this information in a common framework and restricting ourselves
to the three known neutrinos, we arrive at the scenario with three
almost mass-degenerate neutrinos with the mass differences quoted above
\cite{moha}. Such spectrum could arise, for instance,  from the
imposition of a cyclic permutation symmetry among the generations as
pointed out by Harrison, Perkins and Scott \cite{HPS}.

Neutrinos with masses in this range could decay radiatively provided that
they posses an anomalous transition magnetic moment of the form
\begin{equation}
L_{trans}=\frac{1}{2} \bar\nu^\prime \sigma_{\alpha\beta} (\mu+ d\gamma_5) \nu
F^{\alpha\beta}\; +\; h.c.\; ,
\end{equation}
which would give a lifetime for the $\nu\rightarrow \nu^\prime \gamma$ decay
\begin{equation}
\tau^{-1}=(|a|^2+|b|^2)\frac{(\Delta m^2)^3}{8 \pi m^3}\;\; .
\end{equation}
where $|a|=|\mu|\;\; (2\; \mbox{Im}(\mu))$ and $|b|=|d|\;\;
(2\;\mbox{Re}(d))$ for Dirac (Majorana) neutrinos.

The supernova SN1987 limits on this decay mode \cite{SN1987} only apply  to
neutrinos with non-degenerate masses. For mass-degenerate neutrinos  there
exist limits on this decay mode from a laboratory search for gamma-rays from
decaying  $\bar\nu_e$'s produced at a nuclear reactor
\cite{reactor,reactornew}.

In this letter we study the possibility of imposing stringent limits  on this
transition using the CERN neutrino beam. The CERN neutrino beam produces
predominantly $\nu_\mu$'s with fractions of 6\% $\bar\nu_\mu$'s,
0.7\%  $\nu_e$'s  and 0.2\%  $\bar\nu_e$'s. The mean energy of the beam is
27 GeV.  At present two experiments, NOMAD \cite{nomad} and CHORUS,
\cite{chorus}  search for the appearance of $\nu_\tau$ in the CERN beam due to
neutrino oscillations. They expect a sample of about 1 million $\nu_\mu$
charged current (CC) events and the experiments can reach sensitivities of
few $\times 10^{-4}$ in the  $\nu_\mu\rightarrow \nu_\tau$ channel.

The proposal is to use a
resonant high-quality cavity intercepting the beam line to stimulate the
neutrino conversion by absorption or emission of resonant photons inside the
cavity, much as  the experiment proposed by Matsuki and Yamamoto in
Ref.\cite{Matsuki} for solar or
reactor neutrinos. The large number of photons in the cavity enhances the
conversion and improves substantially the accessible range of lifetimes.

In a realistic vein, we will use as resonant cavity one of the
LEP radio-frequency cavities. These are cylindrical cavities with a diameter of
60 cm and
a length of 20 cm along the beam direction. The transition amplitude for the
conversion is given by
\begin{equation}
T_{fi}=\int d^4 x L_{trans} (x)\; .
\end{equation}
For highly relativistic neutrinos moving along the $z$
direction with helicity  $s,s'=\pm 1$ and momentum $k$,
$k^{\prime}$, the wave  function is given by
\begin{equation}
\nu^{(\prime)} (x)\approx
\frac{\exp(-i k^{(\prime)}\cdot x)}{\sqrt{2 \bar V}}
\left( \begin{array}{c}
\phi_{s^{(\prime)}} \\ s ^{(\prime)}\phi_{s^{(\prime)}} \end{array}\right)
\end{equation}
with  $\phi_1=\left(\begin{array}{c} 1 \\[-0.3cm] 0 \end{array}\right)$
and  $\phi_{-1}=\left(\begin{array}{c} 0\\[-0.3cm] 1 \end{array}\right)$ .
Here, $\bar V$ is the volume in which the neutrino field is quantized.

We are going to consider the first transverse magnetic mode of the  resonant
cavity. The electric field is then aligned with the incident neutrino beam.
This configuration will give the largest conversion rate since it corresponds
to photons polarized in the same direction as the neutrino beam.  The
electromagnetic fields inside the cavity are given by:
\begin{equation}
\begin{array}{l}
{\cal E}_z={\cal{E}}_0 J_0(x_{01} r/R) \exp(- i \alpha w t) \\
{\cal{B}}_\phi=\mp i {\cal{E}}_0 J_1(x_{01} r/R) \exp(-i \alpha w t) \\
\end{array}
\end{equation}
where $r$ is the distance to the cavity axis in the transverse plane and
${\cal{B}}_\phi$ is the azimuthal component of the magnetic field. All other
components are zero. $J_i$ are the
Bessel functions and $x_{01}=2.402$ is the first zero of the $J_0$ function.
$w$ is the characteristic frequency of the cavity $w=x_{01}/R=1.4
\times 10^{-6}$ eV. $\alpha=\pm 1$ corresponds to the
process of photon emission $(-1)$ or absorption $(+1)$.

\begin{figure}
\begin{center}
\mbox{\epsfig{file=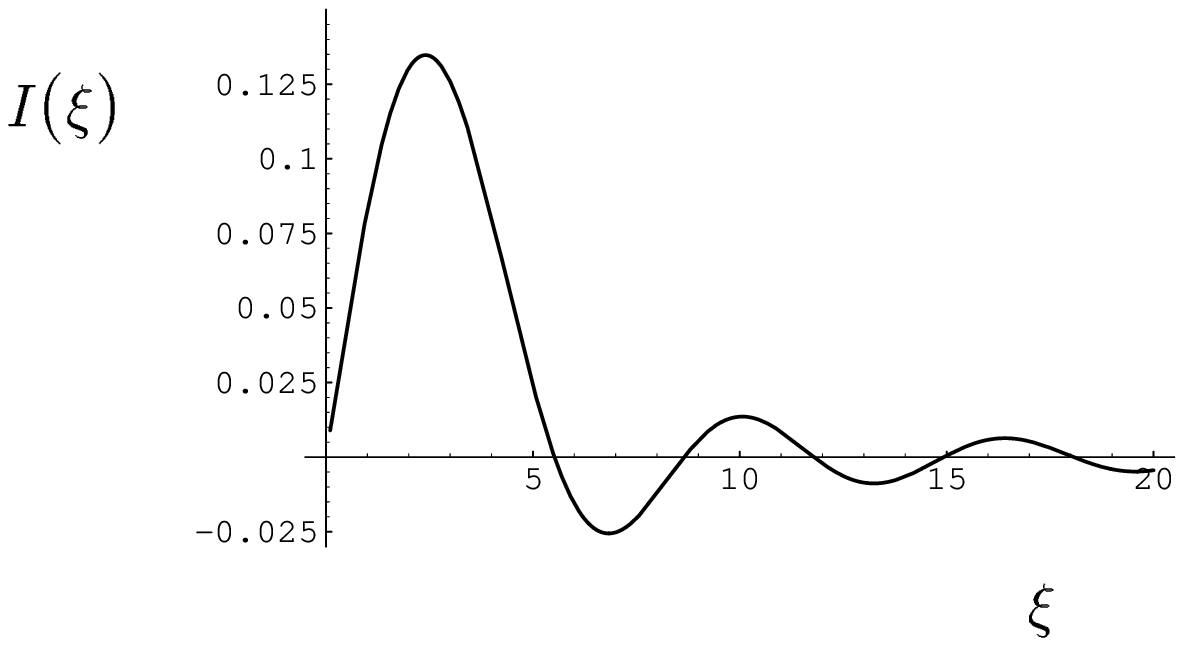,bbllx=60,bblly=200,bburx=600,bbury=600}}
\end{center}
\caption{The form factor $I(\xi)$.}
\label{fig1}
\end{figure}

The transition cross section is
\begin{equation}
\begin{array}{ll}
\sigma =&  \int \frac{\displaystyle d^3 k'}{\displaystyle(2 \pi)^3} \;
\frac{\displaystyle \bar V^2}{\displaystyle T} \;
{\displaystyle\sum_\alpha}\;\left|\frac{1}{2 \bar V}\; (a +b) \;(s'-s)\; \int
d^4x\;[\;i
B_x-s B_y\;] \; \exp(i\Delta k \cdot x)\right|^2 \\
 & =\frac{\displaystyle 4 {\cal E}_0^2}{\displaystyle \pi^2}
\;|a + b|^2 \;V^2 \;{\displaystyle\sum_\alpha}
\int d^3 k' \;\delta (E'-E-\alpha w) \;
\frac{\displaystyle \sin^2(\beta/2)}{\displaystyle\beta^2} \;I^2 (\xi)\\
& = 8 \;{\cal{E}}_0^2 \;|a +b|^2  \;L \;{\displaystyle\sum_\alpha}
\int d\xi \;\xi \;\frac{\displaystyle E+\alpha w}{\displaystyle k'_z}\;
\frac{\displaystyle \sin^2(\beta/2)}{\displaystyle \beta^2} \;I^2 (\xi)
\end{array}
\label{cross}
\end{equation}
where $|a+b|=|\mu-d|\;\; (2|\mbox{Im}(\mu)-\mbox{Re}(d)|$) for incident
left-handed Dirac
(Majorana) neutrinos.
$V=\pi R^2 L$, $\beta=L(k'_z -k_z)$, and  $\xi=k'_T R=
\sqrt{({k'}_x^2+{k'}_y^2)}R$, and the  function $I(\xi)$
\begin{equation}
I(\xi)=\int_0^1 d\rho \rho J_1(x_{01} \rho) J_1(\xi \rho)
\end{equation}
is non-negligible for $\xi\lesssim {\cal O}(10)$, as seen if Fig. \ref{fig1}.
The last step in Eq.(\ref{cross}) is obtained from integration over the
azimuthal angle and over $dk'_z$, this last one using the delta of energy
condition.

The interpretation of Eq.(\ref{cross}) is the following. In the first line we
see that the transition amplitude vanishes unless the helicity is flipped in
the conversion ($ss'=-1$), as expected from a magnetic transition.  The
$\delta$-function in the second line represents energy conservation  resulting
from integration over a large time interval $T\sim L/c \gg \hbar/E$. On the
other
hand since the space integral is not performed over an infinite volume, the
photon momentum is not fixed. If the integral was performed over an infinite
volume the functions verify  ${\displaystyle
\lim_{V\rightarrow\infty}} V\sin(\beta/2)/\beta I(\xi)\rightarrow \delta
(\Delta k_z) \delta({k'}_T^2 -w^2)$,  {\sl i.e.} the momentum along the
beam direction is conserved since the magnetic field lies in the transverse
plane and the transverse momentum transfer must be equal to the photon
energy. In consequence the transition would only be possible for energies
$E=\frac{|\Delta m^2|}{2 w}$.

Since we are restricted to the finite volume of the cavity, there is a
spread on the photon momenta of the order of the inverse of the spatial
dimension of the cavity  ($\xi\sim {\cal O}(1)$, $\beta\sim {\cal O}(1)$). For
these values  the functions  $I(\xi)$ and $\sin(\beta/2)/\beta$ are not
negligible and the variables verify the energy conservation condition
\begin{equation}
\xi^2=R^2\left(2\alpha E w +\Delta m^2 +w^2 -\frac{\beta^2}{L^2}
-2 \frac{\beta}{L} |k| \right)
\label{delta}
\end{equation}
independently of the value of the initial energy $E$ provided this is much
larger than the photon energy $w$ and that $E w/ \Delta m^2\gg 1$, {\sl i.e.}
neutrinos with any large enough energy to verify these conditions can  absorb
or emit a resonant photon in the cavity.  For values of $\xi$
not large enough for $I(\xi)$ to be negligible, the
condition \ref{delta}  on $\beta$ does not depend on $\xi$,
\begin{equation}
\beta\simeq \alpha Lw\; \Rightarrow\; \Delta k_Z= \alpha w\;.
\end{equation}
Therefore the characteristic energy-momentum transfer in the transition  is
extremely small compared with the incident neutrino energy. In other words, to
a very good approximation the neutrino maintains its energy and momentum during
the transition and the interaction with the magnetic  field translates only in
the change of its flavour and the flip of its helicity. The cross section is
consequently independent of the neutrino energy, provided this is large enough.

Finally we can express the field amplitude ${\cal E}_0$ in terms of the number
of
photons in the cavity using the normalization condition
\begin{equation}
N_\gamma w =\frac{1}{2}\int dV (|{\cal{E}}|^2+|{\cal{B}}|^2)=\frac{1}{2} V
{\cal{E}}_0^2 J_1^2(x_{01}) \left[1+\frac{4}{x_{01}^2}\right]=
0.23 V {\cal E}^2_0
\end{equation}
and we can write the  transition rate as
\begin{equation}
\begin{array}{ll}
R=& \frac{\displaystyle \Delta N_\nu}{\displaystyle N_\nu}
=\frac{\displaystyle \sigma}{\displaystyle \pi R^2}=
\frac{ \displaystyle 16 N_\gamma |a+b|^2}{\displaystyle 0.23 \pi R^2 L w}
B\\[+0.3cm]
 &\simeq 50 \kappa_\gamma B \left(\frac{\displaystyle Q}{\displaystyle
10^9}\right)
\left(\frac{\displaystyle P}{\displaystyle 100\; \mbox{W}}\right)
\left(\frac{\displaystyle m}{\displaystyle \mbox{eV}}\right)^3
\left(\frac{\displaystyle \mbox{eV}^2}{\displaystyle \Delta m^2}\right)^3
\left(\frac{\displaystyle 10^{-6} \;\mbox{eV}}{\displaystyle w}\right)
\left(\frac{\displaystyle 10 \mbox{cm}}{\displaystyle L}\right)
\left(\frac{\displaystyle  \mbox{s}}{\displaystyle \tau}\right)
\end{array}
\label{ratio}
\end{equation}
with $B =\sin^2(Lw/2) \int d\xi \xi I^2(\xi)\simeq 0.5$. In the last line  we
have used the relation between the number of photons, the energy, the quality
factor $Q$, and the power supplied to the cavity $P$,   $N_\gamma=4\times
10^{26}
(Q/10^9) (P/100\;\mbox{W}) (10^{-6}\; \mbox{eV}/w)^2$ and we have  defined
\begin{equation}
\kappa_\gamma= |a +b|^2/(|a|^2+|b|^2) \sim {\cal O}(1)\; .
\end{equation}

We now turn to the estimate of the sensitivity attainable at the CERN
experiments for this transition rate. Notice that the
signature is different for  Majorana or
Dirac neutrinos as a consequence of the helicity  flip
in the transition. If neutrinos are Dirac particles the neutrinos in the
beam will convert into sterile right-handed neutrinos. The signal would
then be the disappearance of  $\nu_\mu$'s from the incident beam.
If, on the other hand neutrinos are Majorana particles,
the incoming neutrino will transform into a right-handed active neutrino,
which will produce positive-charge leptons in its CC interactions with
the detector. The signal will then be  the appearance of $\tau^+$ or
an excess of $e^+$.

For the purpose of illustration we use the NOMAD detector. NOMAD is in
essence an electronic bubble chamber, with a continuous target of alternating
panels of light material and drift chambers, followed by  a transition
radiation detector and an electromagnetic calorimeter. All the  detectors are
located inside a $0.4$ T magnetic field. NOMAD measures essentially all
the charged tracks and photons in the event, allowing for a good
reconstruction of the transverse missing momentum in magnitude and direction.
These features permit to study both the $\tau^+$ and the $e^+$ channels.  The
transverse area of the detector is $S=3\times 3\;$m$^2$ and the aim is to
collect $1.1\times 10^{6}$ $\nu_\mu$ CC interactions in the full detector
volume. The expected fraction  of $\nu_\mu$ CC interactions from neutrinos
intercepted by the cavity is the fraction of solid angle covered
by the cavity. This depends on where the cavity is located, since the
beam has a Gaussian profile with $\sigma$ growing linearly with the distance to
the production point and reaching  $\sigma=1$ m at the detectors. If the cavity
is
set at the point where the beam emerges from the ground at a distance of about
150 m from the detector $\sigma=80\;$cm and
\begin{equation}
N_{ev}^{cavity}=1.1\times 10^6
\frac{ \displaystyle \left(2 \pi \int_0^{0.3} \exp(-r^2/0.8^2) r dr\right)}
{0.8^2 \displaystyle \left(\int_{-1.5}^{1.5}
\exp(-x^2) dx\right)^2 } =1.5 \times 10^{5} \; .
\end{equation}
With this number of events one could reach a sensitivity
on the measurement of the ratio $R$ in Eq.(\ref{ratio})
of the order of $1$\%  for a ``disappearance type''  experiment.
Correspondingly one expects  $9\times 10^{3}$ CC from $\bar\nu_\mu$
interactions which constitute the main source of background for the
$\nu_\mu\rightarrow\bar\nu_\tau$ appearance. For the
$\nu_\mu\rightarrow\bar\nu_e$
transition, one should observe an excess of $e^+$ over the
$3\times 10^{2}$ expected events from the interaction of the  $\bar\nu_e$'s
present in the beam. Comparing the interactions observed with the RF cavity
on and off, a sensitivity $R\simeq 10^{-3}$ can be reached for the
$e^+$ appearance experiments. It may be possible to push the sensitivity down
to $R\simeq 10^{-4}$ for the $\bar\nu_\tau$ appearance channel.

\begin{figure}
\begin{center}
\mbox{\epsfig{file=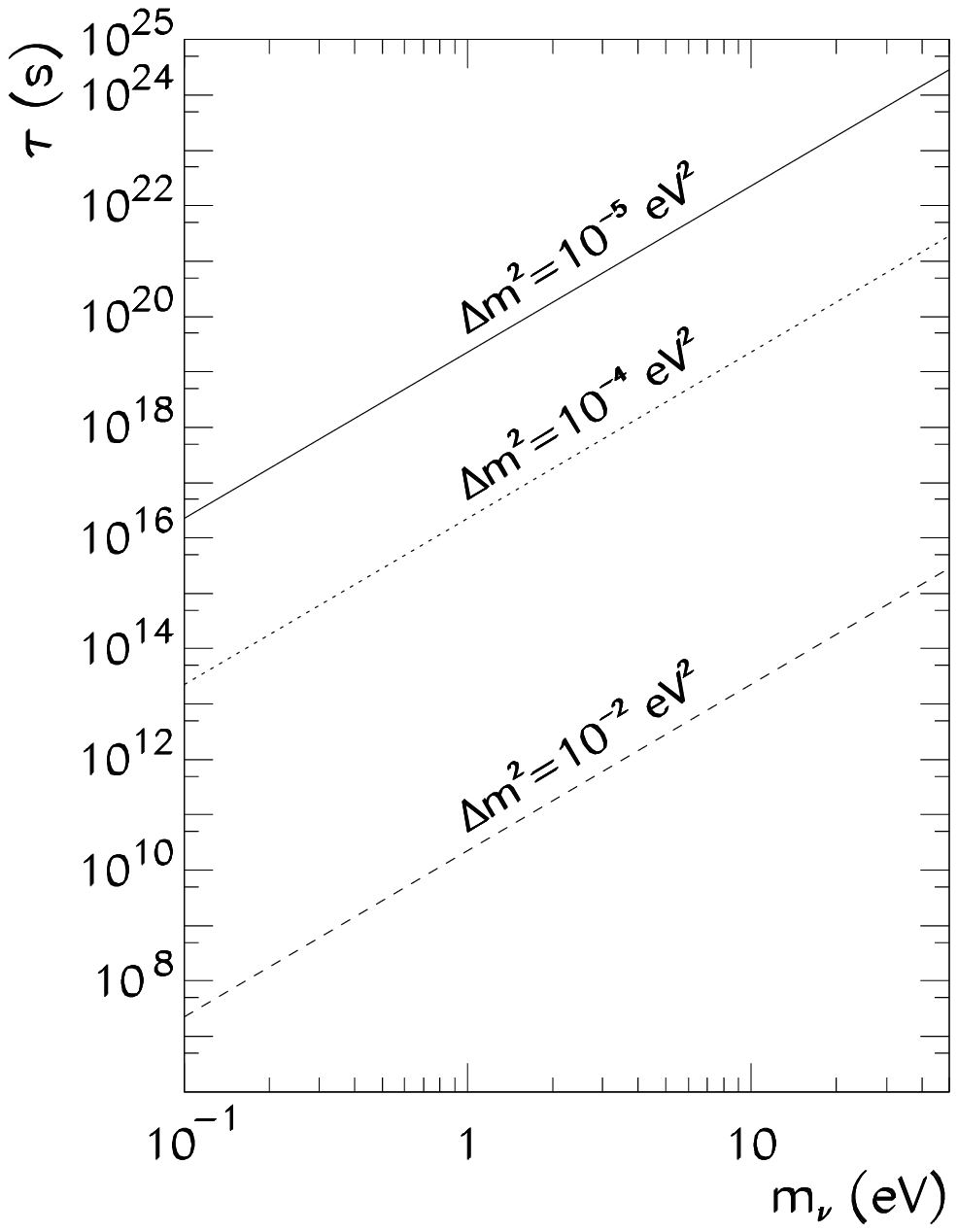,bbllx=60,bblly=200,bburx=600,bbury=600}}
\end{center}
\caption{Lower limits on the radiative decay lifetime attainable
at the CERN experiments by stimulated conversion, assuming a sensitivity
$R=10^{-3}$ as a function
of the neutrino mass for almost mass-degenerate neutrinos with mass
squared differences in the range 10$^{-2}$--$10^{-5}$.}
\label{fig2}
\end{figure}

Such an experiment would impose a limit on the anomalous transition magnetic
moment. To illustrate this, take $d\approx -\mu$. Then
\begin{equation}
|\mu|^2\lesssim 2 \times (10^{-5} \mbox{--} 10^{-7})\; \mu_B^2
\end{equation}
for $R\lesssim 10^{-2}$--$10^{-4}$.
Assuming that there is no cancellation between $d$ and $\mu$,  {\it i.e.}
$\kappa_\gamma \sim {\cal O} (1)$ the experiment can also  impose a lower
limit on the lifetime of the neutrino radiative decay.
In Fig. \ref{fig2} we show the attainable lower limit on the decay lifetime
for an applied power
$P=$100 W and a quality factor $ Q=2.5 \times 10^9$, characteristic of the LEP
cavities. The limit is shown   as a function of the neutrino mass for
mass-squared differences in the interesting range for the solar and atmospheric
neutrino problem. As can be seen in the figure, the attainable limits are
comparable to the age of the Universe, $\tau=10^{17}$ s,  for
$\Delta m^2/m \lesssim  6\times 10^{-5}$ eV.

Our calculation is also valid for the case $\Delta m^2 \ll m$, for which we
get a limit
\begin{equation}
\left(\frac{\tau}{s}\right) \left(\frac{m}{eV}\right)^3 \gtrsim
2\times (10^3\mbox{--}10^{5}) \; ,
\end{equation}
much weaker than the existing limits from SN1987A data \cite{SN1987}
$(\tau/s) (\mbox{eV}/m)> 1.7\times 10^{15}$, in the interesting mass range. The
supernova limit is based on the non observation  of an energetic gamma-ray
burst produced by the neutrino decay. For almost-degenerate neutrinos the
photon energy is suppressed with respect to the non-degenerate case, by a
factor
$\Delta m^2/m^2$ making the limit uninteresting for the parameter range
indicated by solar, atmospheric and dark-matter data.

Limits on radiative decay are also available from reactor data
\cite{reactor,reactornew}. The limit in Ref. \cite{reactor}  for non-degenerate
neutrinos is  $(\tau/\mbox{s}) (\mbox{eV}/m)> 3.8\times 10 $ clearly weaker
than the supernova limit. For mass-degenerate neutrinos the limit is still
valid but now it takes the form $(\tau/\mbox{s}) (\mbox{eV}/m)\gtrsim 3.8\times
10 (m^2/\Delta m^2)^{-a}$,  where $(-a)$ is the slope of the neutrino spectrum
of the reactor, $a=3$--$4$ \cite{reactor2}, becoming also uninteresting. In
Ref. \cite{reactornew} a limit is imposed for  almost-degenerate neutrinos of
the order  $(\tau/\mbox{s}) (\mbox{eV}/m)\gtrsim 10^{-1} F(\Delta m^2/m^2)$
where $F(\Delta m^2/m^2)\leq 1$ in the range  $10^{-7}\leq \Delta m/m\leq
10^{-1}$ and $F$ vanishes out of this range. This constraint is much weaker
than the attainable limit with the experiment proposed here. Notice also that
reactor limits only apply to $\nu_e$ radiative decays.

We conclude that an experiment searching for anomalous magnetic  neutrino
transitions at the CERN beam, stimulated with the use of a LEP radio-frequency
cavity would dramatically improve the existing limits on  neutrino  transition
magnetic moments. The characteristic signature would be the appearance of
$\tau^+$ or an excess of $e^+$ if the neutrinos are Majorana particles. If
neutrinos are Dirac particles one should expect a decrease in the number of
detected $\mu^-$. Comparing the data obtained with and without the cavity, it
is
possible to impose severe limits on the corresponding radiative decay  lifetime
for neutrinos with masses in the interesting cosmological range and mass
differences compatible with a solution  to  the solar and atmospheric neutrino
puzzles.
\newpage
\begin{center}
{\bf \large ACKNOWLEDGEMENTS}
\end{center}
The authors wish to thank J.J. Gomez-Cadenas, R. Vazquez and A. de Rujula
for useful discussions and careful reading of the manuscript.

\newpage

%
\protect

\begin{references}
%
\bibitem{solar} GALLEX Collaboration, Phys.\ Lett.\ {\bf B327} (1994) 377;
SAGE Collaboration, Phys.\ Lett.\ {\bf B328} (1994) 234; Homestake
Collaboration,  Nucl.\ Phys.\ {\bf B38} (Proc. Suppl.)  (1995) 47; Kamiokande
Collaboration, {\sl ibid.}, 55.
%
\bibitem{atmos} Kamiokande Collaboration, H.\ S.\ Hirata {\sl et al.},
Phys.\ Lett.\ {\bf B280}  (1992) 146; Y.\ Fukuda {\sl et al.},  Phys.\
Lett.\ {\bf B335}  (1994) 237; IMB Collaboration, D.\ Casper {\sl et al.},
Phys.\ Rev.\ {\bf D46} (1992) 3720.
%
\bibitem{dark1} E.\ L.\ Wright {\sl et al.}, Astrophys.\ J.\ {\bf 396}
(1992) L13; M.\ Davis, F.\ J.\ Summers and D.\ Schlegel, Nature {\bf 359}
(1992) 293; A.\ N.\ Taylor and R.\ Rowan-Robinson, {\sl ibid.}, 396;
J.\ A.\ Holtzman, and J.\ R.\ Primack, Astrophys.\ J.\ {\bf 405} (1993) 428.
%
\bibitem{dark2} J.\ R.\ Primack, J.\
Holtzman, A.\ Klypin and D.\ O.\  Caldwell, UC Santa Cruz preprint SCIPP 94/28
and references therein.
%
\bibitem{moha} D.\ O.\ Caldwell
and R.\ N.\ Mohapatra, Phys.\ Rev.\ {\bf D48} (1993) 3259.
%
\bibitem{HPS} P.H. Harrison, D.H. Perkins and W.G. Scott, Phys. Lett. {\bf
B349} (1995) 137.

\bibitem{SN1987}F. von Feilitzsch and L. Oberauer, Phys.\ Lett.\  {\bf B200}
(1988) 580; E. W. Kolb and M.S. Turner, Phys.\ Rev.\ Lett.\ {\bf 62} (1989)
509;  M.S. A. Bershady, M. T. Ressel and M.S. Turner, Phys. Rev. Lett. {\bf 66}
(1991) 1389.
%
\bibitem{reactor} L. Oberauer, F. von Feilitzsch and R.L. M\"ossbauer, Phys.
Lett. {\bf B198} (1987) 113.
\bibitem{reactornew} J. Bouchez, B. Pichard, J.P. Soirat and M. Spiro,
Phys. Lett. {\bf B207} (1988) 217; L. Oberauer,  F. von Feilitzsch,
C. Hagner, G. Kempf and R.L. M\"ossbauer, Nucl. Phys. {\bf B} (Proc. Suppl.)
{\bf 28A} (1992) 165.


%
\bibitem{nomad} NOMAD Collaboration, P.\ Astier et al.,
CERN-SPSLC/91-21 (1991), CERN-SPSLC/91-48 (1991), SPSLC/P261 Add. 1 (1991).
%
\bibitem{chorus} CHORUS Collaboration, N.\ Armenise et al.,
CERN-SPSC/90-42 (1990).
%
\bibitem{Matsuki} S.\ Matzuki and K. Yamamoto, Phys.\ Lett.\ {\bf B289} (1992)
194.
%
%
\bibitem{reactor2} P. Vogel {\sl et al.}, Phys.\ Rev.\ {\bf C24} (1981) 1543.
%
\end{references}
\end{document}